\documentclass[reprint,aps,prb,superscriptaddress,twocolumn,showpacs]{revtex4-1}
\usepackage[utf8]{inputenc}
\usepackage{graphicx}
\usepackage[pdftex,hypertexnames=false,pdfpagemode=UseThumbs,hidelinks
,linkbordercolor={1 1 1}, citebordercolor={1 1 1}
,pdfstartview=FitH]{hyperref}
\usepackage{amsmath}
\usepackage{wasysym}
\usepackage{bbold}
\DeclareMathAlphabet{\mathpzc}{OT1}{pzc}{m}{it}

\begin{document}
%%%%%%%%%%%%%%%%%%%%%%%%%%%%%%%%%%%%%%%%%%%%%%%%%%%%%%%%%%
%\title{Braiding Localized Mid-Gap States in a Four-Site Model with Noise}
%\title{Complementary Quality Measures for Noisy Braiding Operations in a Minimal Y-Junction}
\title{Complementary Quality Measures for Noisy Braiding Operations}
\author{Matthias Droth}
\email{matthias.droth@gmail.com}
\affiliation{Laboratoire de Physique, \'{E}cole Normale Sup\'{e}rieure de Lyon, 69007 Lyon, France}
%%%%%%%%%%%%%%%%%%%%%%%%%%%%%%%%%%%%%%%%%%%%%%%%%%%%%%%%%%
\begin{abstract}
Topological quantum computing with non-abelian anyons in a network of one-dimensional chains relies on braiding operations. In real devices, a noisy environment may compromise these braiding operations. In order to assess the failure acquired during braiding with noisy parameters, I define three quality measures. To keep the results as general as possible, I study these quality measures in a model with minimal assumptions that still allows for different kinds of noise.
\end{abstract}
\maketitle
%%%%%%%%%%%%%%%%%%%%%%%%%%%%%%%%%%%%%%%%%%%%%%%%%%%%%%%%%%
\section{Introduction}
Topological quantum computing bears the promise of quantum computing with intrinsic fault tolerance against local perturbations due to nonlocal encoding of quantum information [\onlinecite{Feynman1982,MikeIke,Freedman2003,Nayak2008,Willett2013}]. The proposal to perform topologically protected quantum operations by braiding localized non-abelian anyons around each other in a network composed of one-dimensional chains has received wide interest [\onlinecite{Kitaev2001,Alicea2011,Das2012,Beenakker2013,NadjPerge2014}]. These braiding operations can be performed by locally tuning the model parameters such that localized states move spatially along the network and around each other. However, a noisy environment will affect the model parameters and may thus compromise braiding [\onlinecite{Alicea2011,Knapp2016,Sekania2017,Blanter2000,Deblock2003}]. Therefore, measures that quantify the quality of braiding operations in a noisy environment are required.
\begin{figure}[b!]\centering\includegraphics[width=0.48\textwidth]{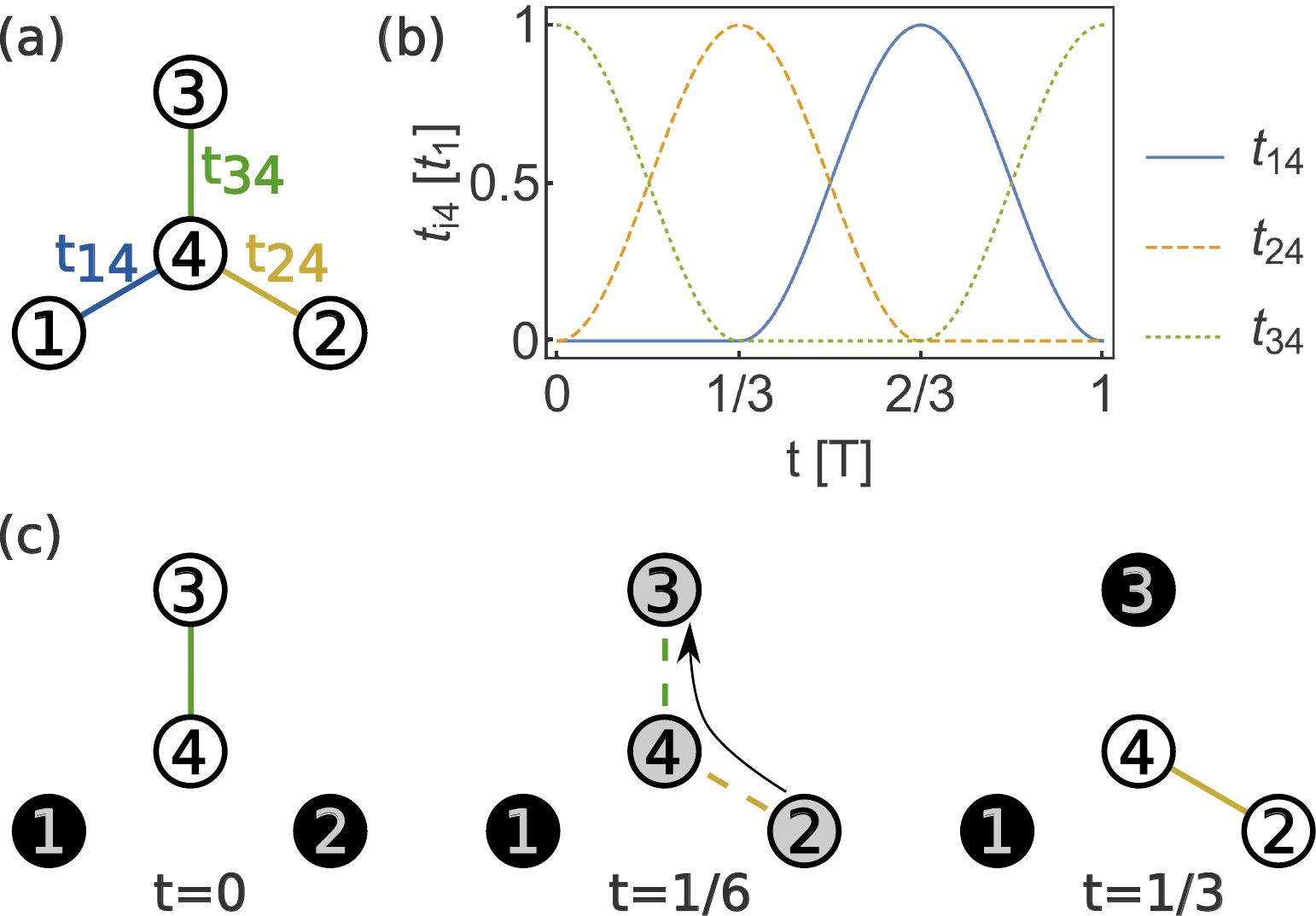}\caption{(Color online) Sketch of the system and of the braiding operation. (a) The system consists of three mutually disconnected outer sites ($i\in\{1,2,3\}$), each of which is only coupled to a central site (4) via a time-dependent hopping $t_{i4}(t)$. These hoppings and the on-site potentials (not shown) may be subject to noise. (b) For $t\in[0,T]$, the value of $t_{i4}(t)$ varies between $0$ and $t_1$ such that localized states are exchanged between the outer sites 1, 2, and 3. (c) Snapshots at times $t=0$, $1/6$, and $1/3$ illustrate how a localized state is moved from site 2 to site 3 as the hoppings $t_{i4}$ evolve as shown in (b).  A darkened site indicates occupation of this site by a mid-gap state.}\label{pic1}\end{figure}

In this article, I use three complementary quality measures to assess the quality of braiding procedures. The \emph{fidelity loss} $\Delta_F$ and the \emph{phase error} $\Delta_{\alpha}$ are calculated after particle exchange and show qualitatively different behavior under different types of noise. The \emph{overlap minimum} $\mathcal{M}(t)$ can be calculated at any time through the braiding operation and can indicate the transition of a localized state from one chain of the network to another.

To study how different noises affect these quality measures, I use the presumably most simple model that supports the braiding of localized mid-gap states, Fig.~\ref{pic1} (a). The model consists of three outer sites, each of which can only be connected to the fourth, central site, thus constituting a minimal chain [\onlinecite{Su1980}]. All three chains share the central site and thus form a Y-junction. By tuning the couplings on and off, a localized state that is isolated on a disconnected outer site of a chain, can be shuttled to the outer site of another chain. This way, two localized states isolated on different outer sites can be exchanged, Fig.~\ref{pic1} (b, c). Further details on the model and the noise it is subject to are the topic of Sec.~\ref{sec2}. The braiding procedure for exchanging two localized mid-gap states is discussed in Sec.~\ref{sec3}. The quality measures for noisy braiding are also defined there. Then, the results are discussed (Sec.~\ref{sec4}) and conclusions are drawn (Sec.~\ref{sec5}).

%%%%%%%%%%%%%%%%%%%%%%%%%%%%%%%%%%%%%%%%%%%%%%%%%%%%%%%%%%
\section{Model, Noise, and Chiral Symmetry}\label{sec2}
The Hamiltonian describing the four-site model is
\begin{eqnarray}
\mathcal{H}=a
\begin{pmatrix}
\mu_1&0&0&t_{14}\\
0&\mu_2&0&t_{24}\\
0&0&\mu_3&t_{34}\\
t_{14}&t_{24}&t_{34}&\mu_4
\end{pmatrix}
\,,\label{hamil}
\end{eqnarray}
with $a\in\mathbb{R}$. In the absence of noise, the on-site potentials $\mu_j$ ($j\in\{1,2,3,4\}$) are zero and the hopping amplitude $t_{14}(t)$ ramps up and down according to a $\sin^2(t)$ dependence [\onlinecite{Sekania2017}],
\begin{eqnarray}
\hspace{-5mm}\frac{t_{14}(t)}{t_1}\!=\!\theta\!\left(\text{mod}(\frac{t}{T},1)\!-\!\frac{1}{3}\right)\sin^2\!\left(\frac{\pi}{2}3\,\text{mod}(\frac{t}{T},1)\!-\!1)\!\right),\label{teq}
\end{eqnarray}
where $t_1$ is the energy unit, $\theta$ is the Heaviside function, and $T$ is the time period. With further time-dependencies $t_{24}(t)=t_{14}(t+T/3)$ and $t_{34}(t)=t_{14}(t-T/3)$, braiding as shown in Fig.~\ref{pic1} is realized.

Noise can be taken into account by adding --- at every specific time step in the discretized braiding operation --- random numbers to each of the default parameters in Eq.~(\ref{hamil}),
\begin{eqnarray}
\mu_j&\to&\tilde{\mu}_j=\mu_j+\Delta\mu\,n^{(j)}_{0,1/3}(t)\,,\\
t_{i4}(t)&\to&\tilde{t}_{i4}(t)=t_{i4}(t)+\Delta t\,n^{(i)}_{0,1/3}(t)\,.\label{subt}
\end{eqnarray}
Here, $\Delta\mu$ and $\Delta t$ denote the strengths of the specific noises. All $\tilde{\mu}_j$ share the same noise strengths, as do all $\tilde{t}_{i4}$. The quantities $n^{(i,j)}_{0,1/3}(t)$ are random numbers drawn from a normal distribution with mean $0$ and standard deviation $1/3$, so that on average less than $0.3\%$ of the added random numbers exceed the specified noise strengths [\onlinecite{statisticsbook}].

\begin{figure}[t!]\centering\includegraphics[width=0.48\textwidth]{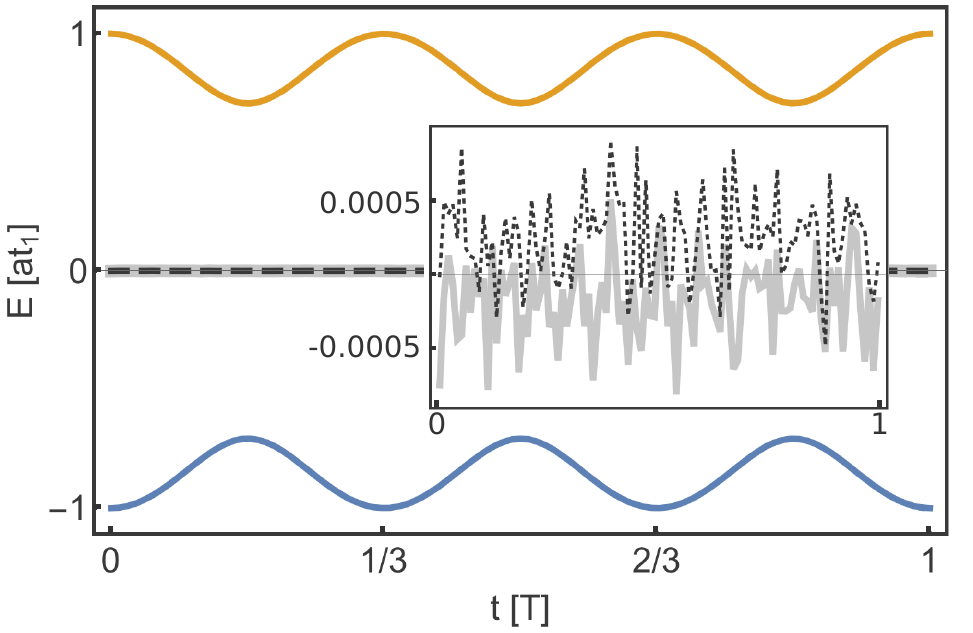}\caption{(Color online) Energy spectrum of the four-site model with noise strengths $\Delta t=\Delta\mu=10^{-3}t_1$. The bonding (blue) and antibonding (yellow) energy levels as well as the energy gap between them are all of the order of $\pm t_1$. In the absence of on-site noise, chiral symmetry implies a symmetric spectrum and mid-gap states at zero energy. The inset confirms the deviation from such a situation due to on-site noise. Pure hopping noise without on-site noise would manifest itself in antisymmetric fluctuations of the bonding and antibonding energies.}\label{pic2}\end{figure}
Assuming $\Delta\mu,\,\Delta t\ll t_1$, the energy difference between mid-gap states and the rest of the spectrum exceeds $0.7a t_1$ at any time. However, adiabatic time evolution demands that this energy difference be much larger than the energy (or frequency\cite{footnote1} $\omega$) associated with the time evolution [\onlinecite{Xiao2010}]. As is evident from Fig.~\ref{pic1} (b), the Hamiltonian undergoes three transitions until it returns to its original form. With $\omega T/3=2\pi$, I thus demand $0.7a t_1\gtrsim1000\,\omega=6000\pi/T$. So I use $a=2.7\times10^4/(t_1T)$ as a dimensionless prefactor in Eq.~(\ref{hamil}).

Sites 1, 2, and 3 are mutually equivalent and different from site 4. So in the basis $(|1\rangle,|2\rangle,|3\rangle,|4\rangle)$, the chiral symmetry operator $\mathcal{C}$ is zero apart from elements $(1,\,1,\,1,\,-1)$ on the diagonal. If its commutator with the Hamiltonian,
\begin{eqnarray}
\mathcal{H}\mathcal{C}-\mathcal{C}\mathcal{H}=
2\begin{pmatrix}
\tilde{\mu}_1&0&0&0\\
0&\tilde{\mu}_2&0&0\\
0&0&\tilde{\mu}_3&0\\
0&0&0&\tilde{\mu}_4
\end{pmatrix}\,,\label{chiral}
\end{eqnarray}
vanishes then the system exhibits chiral symmetry, thus implying a symmetric spectrum [\onlinecite{andrasbook}]. For $\tilde{\mu}_j=0$, the mid-gap states span an exactly degenerate subspace with eigenenergy 0, irrespective of $\Delta t$. Fig.~\ref{pic3} shows the spectrum for $t\in[0,T]$ and $\Delta t=\Delta\mu=10^{-3}$. With $\Delta t\neq\Delta\mu=0$, the the outer eigenenergies fluctuate equally but with opposite sign.
%%%%%%%%%%%%%%%%%%%%%%%%%%%%%%%%%%%%%%%%%%%%%%%%%%%%%%%%%%
\section{Braiding Procedure and Quality Measures}\label{sec3}
Braiding the localized mid-gap states is achieved by time-evolving them along $N$ discrete time steps $\Delta t=T/N$ according to
\begin{eqnarray}
|\psi(0)\rangle\to|\tilde{\psi}(T)\rangle=\left(\mathcal{T}\prod_{j=0}^{N-1}e^{-i\mathcal{H}(j\Delta t)\Delta t}\right)|\psi(0)\rangle\,,\label{Uevo}
\end{eqnarray}
where $\mathcal{T}$ is the time-ordering operator and the Hamiltonian is approximately constant during each step, i.e. $\Delta t\ll1/\omega$. Since $\mathcal{H}(0)=\mathcal{H}(T)$, the instantaneous eigenstate $|\psi(T)\rangle=|\psi(0)\rangle$ is in general different from the time-evolved eigenstate $|\tilde{\psi}(T)\rangle$. Successful braiding requires that the conditions of (i) adiabaticity and (ii) constancy of the Hamiltonian in each time step are sufficiently well fulfilled [\onlinecite{Nayak2008,Sekania2017,Xiao2010}].

Here, the goal is to exchange the localized states such that the state on site 1 (2) at $t=0$ occupies site 2 (1) at $t=T$. To quantify how well this goal is achieved, one may introduce the \emph{overlap matrix} $\mathcal{O}$ that is composed of the overlaps of instantaneous ($\psi$) and time-evolved ($\tilde{\psi}$) eigenstates,
\begin{eqnarray}
\mathcal{O}(t)=\begin{pmatrix}
\langle\psi_1(t)|\tilde{\psi}_1(t)\rangle&\langle\psi_1(t)|\tilde{\psi}_2(t)\rangle\\
\langle\psi_2(t)|\tilde{\psi}_1(t)\rangle&\langle\psi_2(t)|\tilde{\psi}_2(t)\rangle\,.
\end{pmatrix}
\end{eqnarray}
At time $T$, the diagonal elements of $\mathcal{O}$ should vanish and the off-diagonal terms should at best be of unit norm. With $|\cdot|$ being the function that applies the norm to each element of a matrix and $||\cdot||$ returning the norm of an entire matrix, one may define the \emph{fidelity loss} as
\begin{eqnarray}
\Delta_{\rm F}=\frac{\left||\sigma_x-|\mathcal{O}(T)|\,\right|\!|}{2}\in[0,\,1]\,,\label{floss}
\end{eqnarray}
where $\sigma_x$ is the standard Pauli matrix.

While the norm of the off-diagonal terms of $\mathcal{O}(T)$ should be 1, their product should have phase $\alpha\in(0,2\pi]$ as expected upon particle exchange, thus leading ideally to
\begin{eqnarray}
\mathcal{O}_{\rm id}(T)=\begin{pmatrix}
0&e^{i\phi}
\\e^{i(\alpha-\phi)}&0
\end{pmatrix}\,.\label{alphamatrix}
\end{eqnarray}
So the eigenvalues $\pm e^{i\alpha/2}$ of $\mathcal{O}_{\rm id}$ have phases $\alpha/2$ and $\alpha/2-\pi$. With ${\rm arg}\left(\lambda_{1,2}\right)\in(-\pi,+\pi]$ being the phases of eigenvalues $\lambda_{1,2}$ of $\mathcal{O}(T)$, $\tilde{\alpha}=2\,{\rm max}\left({\rm arg}(\lambda_1),\,{\rm arg}(\lambda_2)\right)\in(0,2\pi]$ should ideally be equal to $\alpha$. This leads to the \emph{phase error}
\begin{eqnarray}
\Delta_{\alpha}=\frac{{\rm min}\left(|\tilde{\alpha}-\alpha|,\,2\pi-|\tilde{\alpha}-\alpha|\right)}{\pi}\in[0,\,1]\,.\label{pherror}
\end{eqnarray}

According to the adiabatic approximation, the time-evolved eigenstate should at any time correspond to an instantaneous eigenstate [\onlinecite{Xiao2010}]. Each time-evolved mid-gap state initially corresponds to one of the two localized instantaneous mid-gap states but finally to the respective other, with a transition in between. One may thus calculate the larger overlap $m_i(t)={\rm max}(|\mathcal{O}_{1i}(t)|,\,|\mathcal{O}_{2i}(t)|)$ for each column $i=1,\,2$ of the overlap matrix. The \emph{overlap minimum},
\begin{eqnarray}
\mathcal{M}(t)={\rm min}\left(m_1(t),\,m_2(t)\right)\in[0,\,1]\,,\label{omin}
\end{eqnarray}
is similar to the fidelity loss $\Delta_{\rm F}$ at $t=T$ but in contrast to $\Delta_{\rm F}$, $\mathcal{M}(t)$ can be computed at any time. In the best-case scenario, the two instantaneous mid-gap eigenstates are exactly degenerate and thus span a two-dimensional subspace shared with the time-evolved eigenstates. When both time-evolved eigenstates transition from one instantaneous eigenstate to the respective other, all overlaps and hence $\mathcal{M}(t)$ will have norm $1/\sqrt{2}$.

Next, I evaluate the above quality measures for the model introduced in Sec.~\ref{sec2} and the braiding operation shown in Fig.~\ref{pic1} (b). That is, $\alpha=\pi$, due to fermionic statistics for that model, yet Eqs.~(\ref{floss}-\ref{omin}) apply for any $\alpha$.

%%%%%%%%%%%%%%%%%%%%%%%%%%%%%%%%%%%%%%%%%%%%%%%%%%%%%%%%%%
\section{Results and Discussion of the Braiding Quality Measures}\label{sec4}
Different noise strengths $\Delta\mu$ and $\Delta t$ may affect the quality measures introduced above in different ways. I study this, by specifying the noise strengths $\Delta\mu$, $\Delta t$, and then performing the braiding operation shown in Fig.~\ref{pic1} (b) according to Eq.~(\ref{Uevo}) with $N=10240$ time steps.

\begin{figure}[t!]\centering\includegraphics[width=0.48\textwidth]{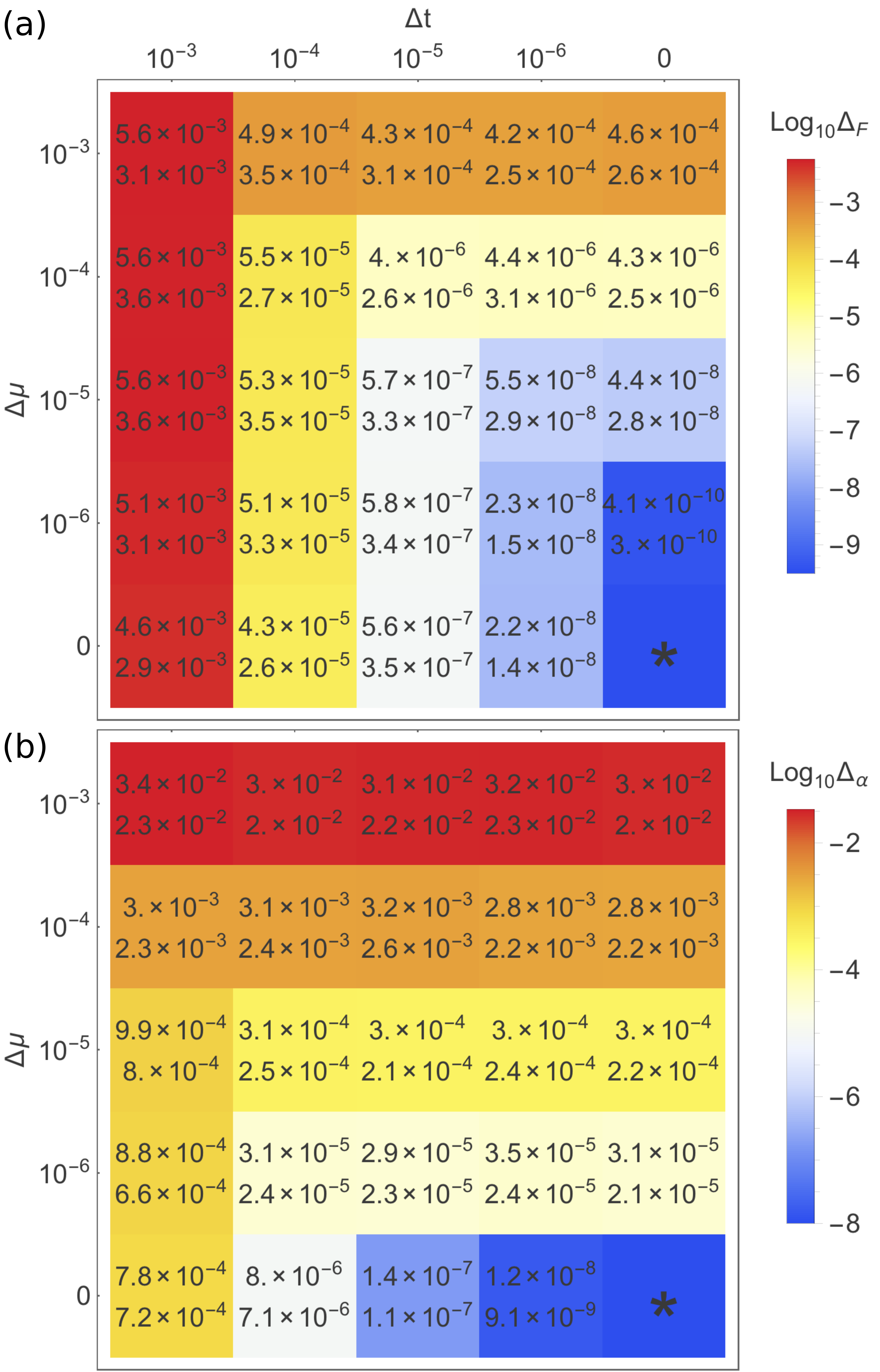}\caption{(Color online) Braiding quality measures. For different noise combinations ($\Delta t$, $\Delta\mu$), I plot in (a) [(b)] the fidelity loss [phase error] as defined in Eq.~(\ref{floss}) [(\ref{pherror})]. The upper [lower] number in each plaquette corresponds to the mean [standard deviation] of the error over 100 independent calculations. The color of each plaquette corresponds to the logarithm with base 10 of the according mean. In the absence of noise (marked by $*$), both errors are always 0 such that their $\log_{10}$ is not defined and their standard deviations vanish.}\label{pic3}\end{figure}

For the fidelity loss, Eq.~(\ref{floss}), and the phase error, Eq.~(\ref{pherror}), I choose $\Delta t$ as well as $\Delta \mu$ (both in units of $t_1$) from the set $\{10^{-3},\,10^{-4},\,10^{-5},\,10^{-6},\,0\}t_1$, thus leading to 25 different noise combinations $(\Delta t,\,\Delta\mu)$. Due to noise being random, the braiding operation is performed 100 times for each noise combination. The results for the fidelity loss $\Delta_F$ [phase error $\Delta_{\alpha}$] are shown in Fig.~\ref{pic3} (a) [Fig.~\ref{pic3} (b)] where each plaquette corresponds to a certain noise combination. The upper [lower] number in a plaquette shows the mean value [standard deviation] of $\Delta_F$ or $\Delta_{\alpha}$ over 100 braiding operations. The background colors of the plaquettes correspond to $\log_{10}\Delta_{F,\alpha}$. Both $\Delta_{F}$ and $\Delta_{\alpha}$ vanish for $\Delta\mu=\Delta t=0$ so their mean values and standard deviations are zero in the according plaquettes. Since $\log_{10}\Delta_{F,\alpha}$ is not defined for $\Delta_{F,\alpha}=0$, the colors of these plaquettes have been assigned manually.

Fig.~\ref{pic3} (a) shows that $\Delta_F$ has a stronger dependence on $\Delta t$ than on $\Delta\mu$. With $\Delta t=10^{-3}$ for example, the magnitude of $\Delta\mu$ seems to affect the fidelity loss no stronger than mere fluctuations around the mean value. Furthermore, $\Delta_F$ varies stronger within rows than within columns. The phase error $\Delta_{\alpha}$ in Fig.~\ref{pic3} (b) behaves in a similar manner but with the roles of $\Delta t$ and $\Delta\mu$ exchanged. Both the fidelity loss and the phase error vanish in the absence of noise and increase as the noise strengths are augmented. At the same time, both measures are affected differently when either the on-site noise $\Delta\mu$ or the hopping noise $\Delta t$ is changed. Therefore, the fidelity loss $\Delta_F$ and the phase error $\Delta_{\alpha}$ are complementary error measures that should both be small.

\begin{figure}[t!]\centering\includegraphics[width=0.48\textwidth]{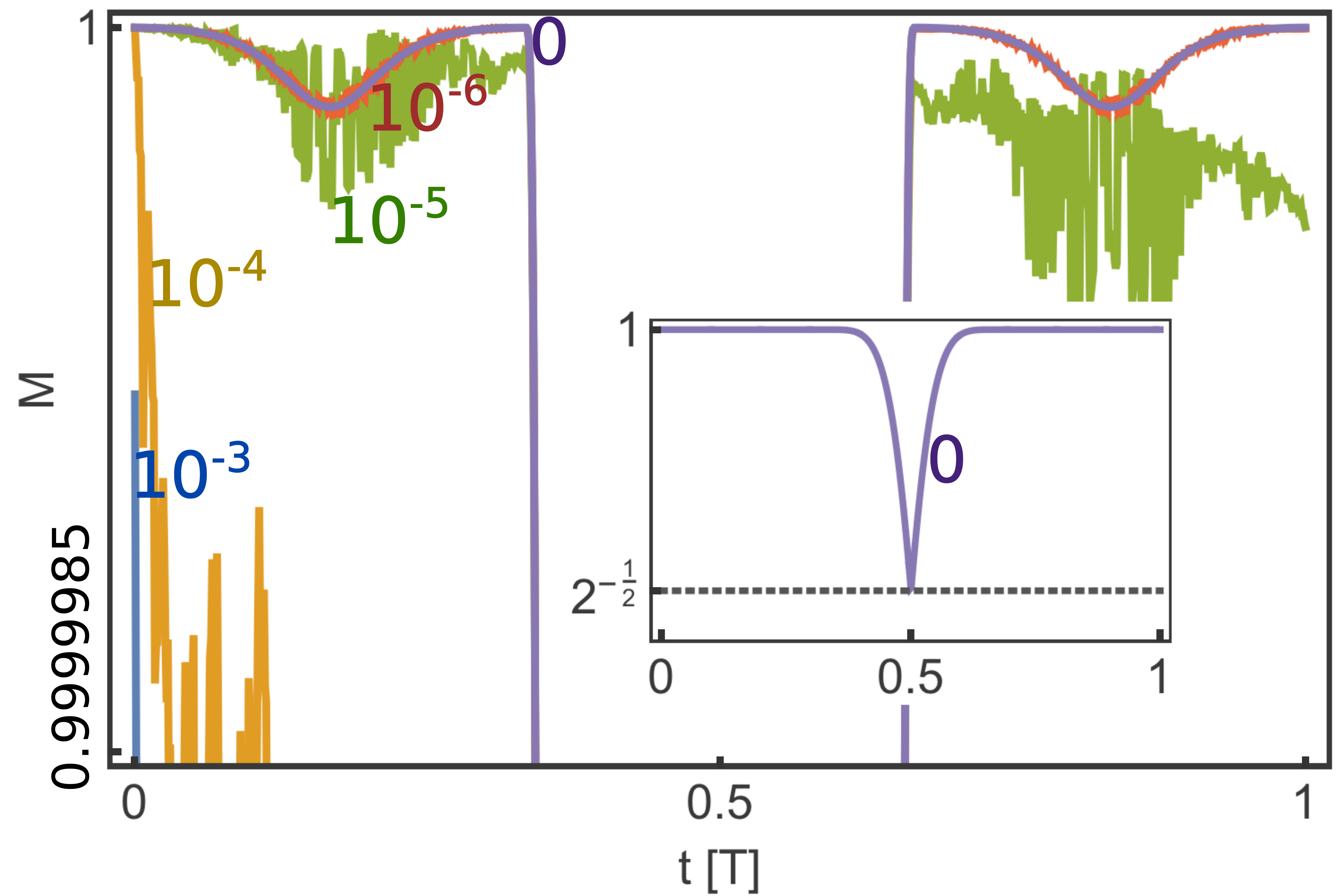}\caption{(Color online) Overlap minimum $\mathcal{M}(t)$ for different noises $\Delta\mu=\Delta t\in\{10^{-3},10^{-4},10^{-5},10^{-6},0\}t_1$, as indicated by the colors. The inset shows that in the absence of noise, the overlap drops to its theoretical minimum value of $2^{-1/2}$, as is expected when the time-evolved eigenstates transition from one instantaneous eigenstate to the respective other. Small depressions at $t=(1/6)T$, $(1/2)T$, and $(5/6)T$ are signatures of a localized state transferring from one to another outer site of the Y-junction.}\label{pic4}\end{figure}

According to Eq.~(\ref{chiral}), chiral symmetry is preserved for $\Delta\mu=0$. Then, the instantaneous mid-gap eigenstates $|\psi_{1,2}(t)\rangle$ span a degenerate two-dimensional zero-energy subspace that is shared by the time-evolved mid-gap states $|\tilde{\psi}_{1,2}(t)\rangle$. During braiding, $|\tilde{\psi}_1(t)\rangle$ and $|\tilde{\psi}_2(t)\rangle$ transition from one instantaneous eigenstate to the respective other while remaining in the 2D-subspace. When these transitions occur simultaneously, the overlap minimum drops to $\mathcal{M}(t)=1/\sqrt{2}$. With $\Delta t=\Delta\mu=0$, this is shown by the purple line in Fig.~\ref{pic4} (main figure and inset). The other lines in Fig.~\ref{pic4} correspond to $\Delta t=\Delta\mu\in\{10^{-6},\,10^{-5},\,10^{-4},\,10^{-3}\}t_1$. Due to the lack of chiral symmetry, the instantaneous mid-gap eigenstates $|\psi_{1,2}(t)\rangle$ are not degenerate in these cases (see also inset of Fig.~\ref{pic2}). As a consequence, a smooth decrease of the overlap minimum to $\mathcal{M}(t)=1/\sqrt{2}$ with subsequent recovery close to 1 is not observed for $\Delta\mu>0$. For $\Delta\mu=0$ and $\Delta t\in\{10^{-6},\,10^{-5},\,10^{-4},\,10^{-3}\}t_1$, $\mathcal{M}(t)$ follows closely the line corresponding to $\Delta t=\Delta\mu=0$ (inset in Fig.~\ref{pic4}) but accumulates a deviation from that line that is of the order of the fidelity loss as shown in the bottom row of Fig.~\ref{pic3} (a).

Apart from the decrease to $1/\sqrt{2}$ for zero noise, Fig.~\ref{pic4} also shows three much smaller depressions centered at $t=(1/6)T,\,(1/2)T$, and $(5/6)T$ that arise due to one of the two localized states changing position: from site 1 to site 3 around $t=(1/6)T$, from site 2 to site 1 around $t=(1/2)T$, and from site 3 to site 2 around $t=(5/6)T$. During these transitions, either $m_1(t)$ or $m_2(t)$ decreases during the first half of the transition and rises again during the second half of the transition. These transitions are visible in Fig.~\ref{pic4} for $\Delta t=\Delta\mu\in\{0,\,10^{-6},\,10^-5\}t_1$ but are masked by the increased loss of fidelity for $\Delta t=\Delta\mu\in\{10^{-4},\,10^{-3}\}t_1$. So for sufficiently low noise, the overlap minimum $\mathcal{M}(t)$ can be used to confirm the transition of a localized state from one chain of the Y-junction to another.

%%%%%%%%%%%%%%%%%%%%%%%%%%%%%%%%%%%%%%%%%%%%%%%%%%%%%%%%%%
\section{Conclusion}\label{sec5}
The results in Sec.~\ref{sec4} confirm that different quality measures are required to properly assess the performance of braiding operations. The fidelity loss $\Delta_F$ and the phase error $\Delta_{\alpha}$ are only evaluated at the end of the braiding operation and are indeed complementary measures that behave differently under different types of noise (Fig.~\ref{pic3}). The overlap minimum $\mathcal{M}(t)$ can be evaluated throughout the braiding operation and displays features linked to the transition of a localized mid-gap state from one chain of the Y-junction / network to another. In Fig.~\ref{pic4}, it also shows sensitivity to the lifting of exact degeneracy of the mid-gap states when chiral symmetry is broken. 

The minimalistic model shown in Fig.~\ref{pic1} (a) and Eq.~(\ref{hamil}) has no parameters other than hoppings $t_{i4}$ and on-site energies $\mu_j$, both subject to noise. So the results presented here should be relevant for any model containing such parameters. Due to the general form of the quality measures in Eqs.~(\ref{floss}-\ref{omin}), this comprises models with non-abelian anyons.
%%%%%%%%%%%%%%%%%%%%%%%%%%%%%%%%%%%%%%%%%%%%%%%%%%%%%%%%%%
\section{Acknowledgements}
I am grateful for discussions with Andr\'{a}s P\'{a}lyi and Titus Neupert as well as for funding provided by the Deutsche Forschungsgemeinschaft (German Research Foundation) within Project No. 317796071.
%%%%%%%%%%%%%%%%%%%%%%%%%%%%%%%%%%%%%%%%%%%%%%%%%%%%%%%%%%

\end{document}